\documentclass[prl,showpacs,twocolumn,amsmath,amssymb]{revtex4}

\usepackage{graphicx}
\usepackage{bm}

\begin{document}

\title{Anisotropic Inverse Cascade toward Zonal Flow in
Magnetically Confined Plasmas}

\author{Alexander M. Balk$^1$ and Peter B. Weichman$^2$}

\affiliation{$^1$Department of Mathematics, University of Utah, Salt
Lake City, UT 84112 \\
$^2$BAE Systems, Advanced Information Technologies, 6 New England
Executive Park, Burlington, MA 01803}




\begin{abstract}

We propose a new mechanism for the generation of zonal flows in
magnetically confined plasmas, complementing previous theories based on
a modulational instability. We derive a new conservation law that
operates in the regime of weakly nonlinear dynamics, and show that it
serves to focus the inverse cascade of turbulent drift wave energy into
zonal flows. This mechanism continues to operate in the absence of the
separation of dynamical scales typically assumed in instability
calculations.

\end{abstract}

\maketitle

Zonal flows refer to a class of highly anisotropic flows that emerge
spontaneously in response to nominally rather weakly anisotropic trends
in the environment. Examples in geophysical fluids include strongly
sheared east-west jets in planetary atmospheres (Jupiter, in
particular). North-south variation of the Coriolis parameter provides
an obvious underlying anisotropy, but is far too weak to directly
explain the observed flow patterns.

Similar flows are believed to exist in magnetically confined
quasi-neutral plasmas, e.g., tokamaks \cite{DIIH2005}, responding in
this case to gradients in the magnetic field, background ion
concentration, plasma temperature, etc. Zonal flows take on added
importance in plasmas because they are believed to provide transport
barriers, leading to the low to high confinement (L-H) transition, and
hence may aid the goal of controlled fusion.

There have been many calculations elucidating conditions under which
small-scale drift wave turbulence can produce a modulational
instability, leading to exponential growth of a zonal flow pattern
\cite{DRHMFS1998,SDS2000,KWPSSS2005,KST2010,CNNQ2010}. The calculations
typically assume a large separation of scales, enabling a simple
description of the growth of an existing zonal flow, pumped by
sufficiently strong resonant small scale fluctuations. Broader
conditions can probably be formulated in terms of the shape of the
drift wave spectrum \cite{CNNQ2010}.

The goal here is to show that energy transfer from small scale
turbulence to large scale zonal flow is a general physical phenomenon
in plasmas, operating irrespective of whether the dominant interactions
are local or nonlocal in scale. We derive a new ``extra'' conservation
law, on top of those for energy and momentum/enstrophy. The inverse
cascade of wave energy follows from standard arguments involving
balance of energy and enstrophy flux in the spectral domain
\cite{Z1992}. The extra invariant places further constraints on the
energy flux, forcing it more and more strongly into the zonal
wavevector sector with increasing scale. This provides a general
mechanism for the observed amplification of zonal anisotropy, and will
be demonstrated quantitatively through analysis of a corresponding
spectral function $\phi_{\bf k}$.

A similar invariant exists in the quasigeostrophic (or CHM \cite{CHM})
equation \cite{BNZ1991,B1991}, and in the shallow water system
\cite{BHW2011} of geophysical fluid dynamics. However, the plasma
system is significantly more complicated, producing, for example, in
addition to the usual CHM ``vector'' nonlinearity, a ``scalar''
nonlinearity \cite{KWPSSS2005,NC1995,OPSSS2004,MHM}. We find it
unlikely, for example, that the generalized Hasegawa-Mima (GHM)
\cite{NC1995} equation (with both nonlinearities) possesses an extra
invariant---this issue will be discussed further below. Given the added
levels of approximation entering such reduced equations, we base our
derivation directly on the more general effective two-dimensional
hydrodynamic equations from which they are derived. We account for
(smooth, large scale) inhomogeneity in the electron temperature (which
leads to the scalar nonlinearity), applied magnetic field $B$, fluid
pressure $P$, and background ion density $n_0$, all on the same
footing. Furthermore, we do not assume any particular common direction
of variation of these parameters (e.g., the ``radial'' direction in a
tokamak \cite{foot:direction}). A single direction $\hat {\bm \gamma}$
emerges naturally, namely the local gradient of the ratio $n_0/B$, with
zonal direction orthogonal to it.

The effective two-dimensional hydrodynamic equations for a magnetically
confined plasma, in the $xy$-plane normal to the applied magnetic field
${\bf B} = B {\bf \hat z}$, are
\begin{eqnarray}
\partial_t {\bf v} + ({\bf v} \cdot \nabla) {\bf v}
+ f {\bf \hat z} \times {\bf v} &=& -\nabla \Phi
\nonumber \\
\partial_t n + \nabla \cdot (n {\bf v}) &=& 0,
\label{1}
\end{eqnarray}
where $n$ is the ion number density and ${\bf v} = (v_x,v_y)$ the
velocity. The ion cyclotron frequency is $f = q_i B/m_i$, where
$q_i,m_i$ are the ion charge and mass. The potential $\Phi = (q_i/m_i)
\varphi + W(n)$ consists an electric potential $\varphi$ (with ${\bf E}
= -\nabla \varphi$), and an ion pressure term ($W$ being the solution
of $m_i n W'(n) = P'(n)$, with pressure $P$ assumed a function of the
density alone). Electrons move rapidly along the magnetic field lines,
and are assumed to be in local equilibrium at a temperature $T$. The
density then follows the Boltzmann distribution, $n = n_0 e^{e
\varphi/T}$ (so $n_0$ is the ion concentration when $\varphi \equiv
0$). The fluctuation contribution to ${\bf B}$ from currents generated
by ${\bf v}$ is assumed small compared to $B$, and is neglected.

We will see that the slow drift wave modes in (\ref{1}) are weakly
coupled to all other motions, and hence support a separate set of
approximate conservation laws. Most importantly, the drift wave
dispersion law, essentially uniquely, admits the new conservation law
\cite{BF1998}.

Irrespective of the relation between $\Phi$ and $n$ (if any), (\ref{1})
leads to convective conservation of potential vorticity
\begin{equation}
(\partial_t + {\bf v} \cdot \nabla) Q = 0,\ \
Q \equiv (\zeta + f)/n,
\label{2}
\end{equation}
where $\zeta = \nabla \times {\bf v} \equiv \partial_x v_y - \partial_y
v_x$ is the vorticity. In the weakly nonlinear limit one may
approximate $Q \approx ({\cal Q} + f)/n_0$, where ${\cal Q} = \zeta - f
h$ and $h = (n - n_0)/n_0$. From (\ref{1}) one obtains the (exact)
equation of motion
\begin{equation}
\partial_t {\cal Q} + \nabla \cdot ({\cal Q}{\bf v})
= f (1+h) {\bm \gamma} \cdot {\bf v},\ \
{\bm \gamma} \equiv \nabla \ln(n_0/f).
\label{3}
\end{equation}

Equation (\ref{2}) gives rise to the usual infinite hierarchy of
conserved integrals. We derive here an invariant of a different type.
It is quadratic in ${\cal Q}$,
\begin{equation}
I = \frac{1}{2} \int d^2r_1 \int d^2r_2 X({\bf r}_1,{\bf r}_2)
{\cal Q}({\bf r}_1,t) {\cal Q}({\bf r}_2,t),
\label{4}
\end{equation}
with some (symmetric) kernel $X$ to be determined. Drift wave energy
and momentum may be expressed this way, but for the plasma system
(\ref{1}) there is an extra choice.

We utilize two small parameters. First, we assume that $n_0, f, T$ vary
slowly, namely on the scale $1/\gamma$, which itself is much larger
than dominant scale $L$ of variations of the fields ${\mathcal Q},{\bf
v}$. The quantity $\mu = \gamma L$ is the \emph{small inhomogeneity}
parameter. The quantities $|\nabla n_0|/n_0, |\nabla f|/f, |\nabla
T|/T, |\nabla {\bm \gamma}|/\gamma$ are all $O(\mu/L)$. This is similar
to the usual beta-plane approximation in geophysical fluid dynamics,
and will allow us, at a critical stage, to perform a local Fourier
analysis. Second, weak nonlinearity constrains the dimensionless
characteristic field amplitude $A \sim {\cal Q}/f \sim {\bf v}/fL$. The
divergence term in (\ref{3}) should be much smaller than the linear
term on the right hand side, leading to the \emph{small nonlinearity}
parameter $\epsilon = A/\mu$. More generally, for a complex turbulent
state, this parameter should be small on all length scales, not just
the dominant scale $L$.

Using typical D-T fusion plasma parameters, $T \sim 10$ KeV, $B \sim 5$
T one obtains Larmor radius $\rho \equiv \sqrt{T/f^2 m_i} \sim 3$ mm.
The drift velocity is $v \approx E/B \sim \varphi/LB$, where $\varphi
\approx T h/e$ is estimated from the Boltzmann relation, yielding
$\zeta/f \sim h (\rho/L)^2$, ${\cal Q}/f \sim [1 + (\rho/L)^2]h$, and
hence $\epsilon \sim Q/f L\gamma \sim [1 + (\rho/L)^2]h/\mu$. Using
inhomogeneity scale $1/\gamma \sim 1$ m, density fluctuation scale $h
\sim 10^{-2}$, and zonal flow scale $L \sim 10$ cm, one therefore
obtains $\mu,\epsilon$ both of order $10^{-1}$. These are indeed small,
well within the range of validity of the theory to follow.

We will prove that there are only three independent choices of the
kernel $X$, for which $I$ is approximately conserved, i.e., may be
considered constant over very long time scales (made more precise
below). The simplest way to do so would be to bound $|\dot I|/I$.
Unfortunately, $I$ contains oscillatory terms that have small
amplitude, but whose time derivatives do not. Taking ${\bf v},{\cal Q}$
as the independent fields, we therefore consider a supplemented
\cite{ZS1988} invariant
\begin{eqnarray}
{\cal I} &=& I
+ \int d_{12} {\cal Q}_1 {\bf F}_{12} \cdot {\bf v}_2
+ \frac{1}{2} \int d_{123}{\cal Q}_1 {\cal Q}_2
{\bf M}_{123} \cdot {\bf v}_3
\nonumber \\
&&+\ \frac{1}{6} \int d_{123}
Y_{123} {\cal Q}_1 {\cal Q}_2 {\cal Q}_3,
\label{5}
\end{eqnarray}
in which, to condense the notation, numerical subscripts stand for the
argument: $d_{12} = d^2r_1 d^2r_2$, ${\bf v}_2 = {\bf v}({\bf r}_2,t)$,
${\bf M}_{123} = {\bf M}({\bf r}_1,{\bf r}_2,{\bf r}_3)$, etc. For
small $\mu,\epsilon$ the added terms will be found to be much smaller
than $I$ but their time derivatives generally are not, and the kernels
$X,{\bf F},{\bf M},Y$ will be determined by demanding that $|\dot {\cal
I}|/f{\cal I}$ be small \cite{foot:Icorr}. The kernel ${\bf M}$ is
symmetric in its first two arguments, and $Y$ is symmetric in all
three. Other cubic terms are possible, involving different combinations
of ${\cal Q},{\bf v}$, but, due to the structure of (\ref{1}), turn out
not to contribute, so we drop them from the outset.

We will see that ${\bf F}_{12} = O(\mu)$, hence to compute $\dot {\cal
I}$ it suffices to approximate (\ref{1}), (\ref{3}) by
\begin{equation}
{\partial_t{\bf v}} = -f {\bf \hat z} \times {\bf v} - \nabla \Phi,\ \
{\partial_t {\mathcal Q}} = f {\bm \gamma} \cdot {\bf v}
- \nabla\cdot({\mathcal Q}{\bf v}).
\label{6}
\end{equation}
Here $\Phi$ depends only on the local density, and is taken to vanish
for the steady state plasma. It suffices as well to use its linearized
form
\begin{equation}
\Phi = {\cal T} h = ({\cal T}/f)(\zeta - {\cal Q})
\label{7}
\end{equation}
with the slow function ${\cal T}({\bf r}) = T({\bf r})/m_i$ in the
standard cold ion limit where one neglects the pressure term.

We define for convenience the combinations
\begin{equation}
G_{12} = ({\cal T}_2/f_2)
\nabla_2 \cdot {\bf F}_{12},\ \
K_{123} = ({\cal T}_3/f_3)
\nabla_3 \cdot {\bf M}_{123}.
\label{8}
\end{equation}
Using (\ref{1}) and (\ref{3}), integrating by parts where necessary to
remove spatial derivatives from the fields, and collecting terms, one
then obtains, to requisite order, $\dot {\cal I}$ with the same four
terms as in (\ref{5}), but with corresponding (appropriately symmetric)
kernels,
\begin{eqnarray}
\tilde X_{12} &=& -(G_{12} + G_{21})
\nonumber \\
\tilde {\bf F}_{12} &=& X_{12} f_2 {\bm \gamma}_2
+ {\bf \hat z} \times {\bf F}_{12} f_2
+ \nabla_2 \times G_{12}
\nonumber \\
\tilde {\bf M}_{123} &=& \delta_{23} \nabla_3 X_{13}
+ \delta_{13} \nabla_3 X_{23}
+ {\bf \hat z} \times {\bf M}_{123} f_3
\nonumber \\
&&+\ \nabla_3 \times K_{123}
+ Y_{123} f_3 {\bm \gamma}_3
\nonumber \\
\tilde Y_{123} &=& -(K_{123} + K_{132} + K_{321}),
\label{9}
\end{eqnarray}
where $\delta_{12} = \delta({\bf r}_1-{\bf r}_2)$, $\nabla \times G =
(\partial_y G,-\partial_x G)$, etc. A number of terms of higher order
in $\mu,\epsilon$ have been dropped. These physically represent higher
order nonlinearity, including interactions between drift waves and
other modes contained in (\ref{1}). The vanishing of $\tilde X_{12}$
and $\tilde Y_{123}$ produce the antisymmetry conditions
\begin{eqnarray}
G_{12} + G_{21} = 0,\ \ K_{123} + K_{132} + K_{321} = 0,
\label{10}
\end{eqnarray}
while the vanishing of $\tilde {\bf F}_{12}$ and $\tilde {\bf M}_{12}$
produce
\begin{eqnarray}
{\bf F}_{12} &=& \frac{1}{f_2} \nabla_2 G_{12}
+ ({\bf \hat z} \times {\bm \gamma}_2) X_{12}
\nonumber \\
{\bf M}_{123} &=& \frac{1}{f_3} \nabla_3 K_{123}
+ {\bf \hat z} \times \bigg[{\bm \gamma}_3 Y_{123}
\nonumber \\
&&\left.+ \frac{1}{f_3} \delta_{23} \nabla_3 X_{13}
+ \frac{1}{f_3} \delta_{13} \nabla_3 X_{23} \right].
\label{11}
\end{eqnarray}
Substituting (\ref{11}) into the right hand sides of (\ref{9}), one
obtains the closed equations
\begin{eqnarray}
\hat {\cal L}_2 G_{12} &=&
({\bm \gamma}_2 \times \nabla_2) X_{12}
\nonumber \\
\hat {\cal L}_3 K_{123} &=&
({\bm \gamma}_3 \times \nabla_3) Y_{123}
- \left(\nabla_3 \frac{\delta_{23}}{f_3} \right)
\times (\nabla_3 X_{13})
\nonumber \\
&&-\ \left(\nabla_3 \frac{\delta_{13}}{f_3} \right)
\times (\nabla_3 X_{23}),
\label{12}
\end{eqnarray}
in which we have defined the operator
\begin{equation}
\hat {\cal L} = -\nabla \cdot \left(\frac{1}{f} \nabla \right)
+ \frac{f}{\cal T}.
\label{13}
\end{equation}

We may write formally $G_{12} = \hat {\cal K}_2 X_{12}$, where $\hat
{\cal K} = \hat {\cal L}^{-1} ({\bm \gamma} \times \nabla)$. Since
$\hat {\cal K}_2$ depends only on ${\bf r}_2$, it does not have any
particular symmetry under interchange of ${\bf r}_1$ and ${\bf r}_2$.
Therefore, given that $X_{12} = X_{21}$ is symmetric, it is generally
impossible to enforce antisymmetry of $G_{12}$. However, to leading
order, one may ignore the ${\bf r}$ dependence of all the parameters,
replacing them by constant characteristic values. In this case $\hat
{\cal K}$ is translation invariant, and one seeks translation invariant
solutions $X_{12} = X({\bf r}_1-{\bf r}_2)$, with $X({\bf r})$ even.
Hence $G_{12} = G({\bf r}_1-{\bf r}_2)$ is odd, and the first of
equations (\ref{10}) is automatically satisfied.

Similarly, given the symmetries of $X_{12}$ and $Y_{123}$, it is
impossible to enforce antisymmetry of $K_{123}$ beyond leading order.
However, again replacing all parameters by constants, a consistent
solution for $Y_{123}$ does exist. It is most conveniently expressed in
Fourier space, where, in particular $f \hat {\cal K} \to i \Omega({\bf
k})$, in which
\begin{equation}
\Omega = \frac{f {\bm \gamma} \times {\bf k}}{k^2+\alpha^2},\ \
\alpha^2 \equiv \rho^{-2} = \frac{f^2}{\cal T}
\label{14}
\end{equation}
exhibits the usual drift wave dispersion relation
\cite{foot:driftwave}. One obtains:
\begin{eqnarray}
\hat Y_{123} &=& \frac{2 A_{123}
\hat \delta_{123}}{i(\Omega_1+\Omega_2+\Omega_3)}
\left[\left(\frac{1}{k_3^2+\alpha^2}
- \frac{1}{k_2^2+\alpha^2} \right) \hat X_1 \right.
\nonumber \\
&&+\ \left(\frac{1}{k_1^2+\alpha^2}
- \frac{1}{k_3^2+\alpha^2} \right) \hat X_2
\nonumber \\
&&\left. +\ \left(\frac{1}{k_2^2+\alpha^2}
- \frac{1}{k_1^2+\alpha^2} \right) \hat X_3 \right],
\label{15}
\end{eqnarray}
with $\hat \delta_{123} = (2\pi)^2 \delta({\bf k}_1 + {\bf k}_2 + {\bf
k}_3)$ reflecting translation invariance, and $A_{123} = \frac{1}{2}
{\bf k}_1 \times {\bf k}_2 = \frac{1}{2} {\bf k}_2 \times {\bf k}_3 =
\frac{1}{2} {\bf k}_3 \times {\bf k}_2$ being the area of the resulting
triangle formed by ${\bf k}_1,{\bf k}_2,{\bf k}_3$.

Up to now, $\hat X({\bf k})$ is arbitrary.  However, (\ref{15})
displays a divergence on the ``three-wave resonant'' surface
\begin{equation}
{\bf k}_1 + {\bf k}_2 + {\bf k}_3 = 0,\ \
\Omega({\bf k}_1) + \Omega({\bf k}_2) + \Omega({\bf k}_3) = 0.
\label{16}
\end{equation}
Only if the term in square brackets vanishes on this surface does a
nonsingular $Y_{123}$ emerge, and this places rather stringent
conditions on $X$, that we will now discuss. On this surface, the
vector $f {\bf k}_{123} = {\bf k}_1 \Omega_2 - {\bf k}_2 \Omega_1 =
{\bf k}_2 \Omega_3 - {\bf k}_3 \Omega_2 = {\bf k}_3 \Omega_1 - {\bf
k}_1 \Omega_3$ is also symmetric under cyclic permutation of its
indices. Defining the ``zonal'' and ``radial'' wavenumbers $p = -\hat
{\bm \gamma} \times {\bf k}$, $q = \hat {\bm \gamma} \cdot {\bf k}$,
one may write the term in square brackets as
\begin{eqnarray}
[\ \cdot \ ] &=& \frac{{\bm \gamma} \times {\bf k}_{123}}
{p_1 p_2 p_3} \left[p_1 \hat X_1 + p_2 \hat X_2 + p_3 \hat X_3 \right],
\label{17}
\end{eqnarray}
and from its vanishing one therefore obtains the condition that $p \hat
X_{\bf k}$ also be conserved on the resonance surface.

The set of kernels satisfying this condition has been investigated at
length \cite{BNZ1991,B1991}. In addition to the obvious choices $\hat
X^Z_{\bf k} = 1$ (enstrophy/zonal momentum), $\hat X^E_{\bf k} =
\Omega({\bf k})/\gamma f p = (k^2+\alpha^2)^{-1}$ (energy), the
\emph{extra invariant}
\begin{equation}
\hat X^M_{\bf k} =
\frac{1}{p} \arctan \frac{\alpha (q+p\sqrt{3})}{k^2}
- \frac{1}{p} \arctan \frac{\alpha (q-p\sqrt{3})}{k^2}.
\label{18}
\end{equation}
The existence of the corresponding invariant (\ref{4}) in plasmas is
the fundamental result of this paper. Since solutions to (\ref{10}) do
not exist beyond leading order in $\mu$, neither does extra
conservation. Intuitively, higher order accuracy in $\mu$ must account
for $\nabla T$, but $\hat X^M$ depends only on the one direction ${\bm
\gamma}$ (via definition of $p,q$). The GHM equation \cite{NC1995}
(with comparable vector and scalar nonlinearities) has both gradients,
and it appears impossible to find an extra invariant at all (even if
$\nabla T \parallel {\bm \gamma}$) \cite{foot:GHM}. Only when one
nonlinearity dominates, and correspondingly, one of the gradients can
be disregarded, does an extra invariant emerge. Our results show that a
consequence of (\ref{1}) is that $\nabla T$ is always higher order.

\begin{figure}

\includegraphics[width=\columnwidth]{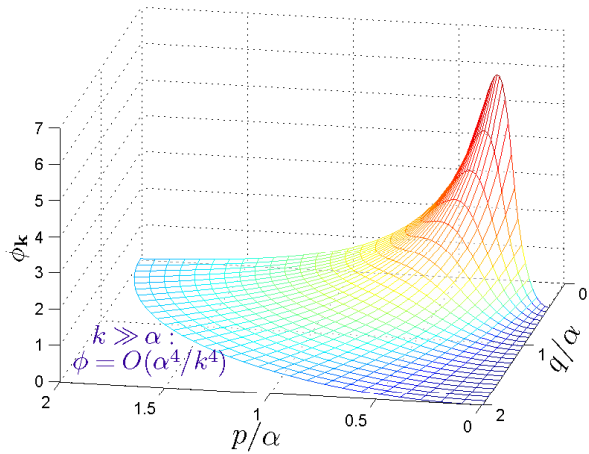}

\caption{3D plot of the ratio (\ref{19}), which measures how much extra
invariant $\tilde I$ is attached to a unit amount of energy with wave
vector ${\bf k}$. This ratio is even in both ``zonal'' wavenumber $p =
-\hat {\bm \gamma} \times {\bf k}$ and ``radial'' wavenumber $q = \hat
{\bm \gamma} \cdot {\bf k}$, with the direction $\hat {\bm \gamma}$
defined in (\ref{3}), and is plotted for $0.4 < k/\alpha < 2$. If one
considers transporting a unit of energy from small scales (large $k$)
toward the origin, one is forced to navigate around the peak, and
squeeze into the valley along the $q$-axis.}

\label{fig:extrainv}
\end{figure}

To understand the significance of (\ref{18}), and its relation to the
formation of zonal flows, consider the linear combination $\tilde I = I
- 2 \sqrt{3} \alpha E = \int \phi_{\bf k} E_{\bf k} d^2k/(2\pi)^2$,
where $E_{\bf k} = \hat X^E_{\bf k}|\hat {\cal Q}_{\bf k}|^2$ is the
energy spectrum, and
\begin{eqnarray}
\phi_{\bf k} &=& \hat X^M_{\bf k}/\alpha
\hat X^E_{\bf k} - 2\sqrt{3}
\label{19} \\
&=& \left\{\begin{array}{ll} \frac{8\sqrt{3} \alpha^4
p^2(q^2+p^2/5)}{k^8} + O[(\alpha/k)^6],
& k/\alpha \gg 1 \\
\frac{8\sqrt{3} \alpha^4 p^2}{q^2(q^2+\alpha^2)^2} + O[(p/q)^4],
& p/q \ll 1.
\end{array} \right.
\nonumber
\end{eqnarray}
Remarkably, $\hat X^E$ and $\hat X^M$ have identical (up to a factor
$2\sqrt{3}\alpha$) asymptotic behavior for large $k$ and for small $p$,
so $\hat \phi \to 0$ when $k \to \infty$ or $p \to 0$. The inverse
cascade follows from the spectral balance needed to maintain
conservation of energy and enstrophy \cite{foot:approx}. As illustrated
in Fig.\ \ref{fig:extrainv} the extra conservation provides additional
constraints leading to zonal flows. Specifically, it follows from
(\ref{19}) that a unit of energy at large $k/\alpha$ carries very small
$\phi$. Therefore transferring energy towards the origin (either by
local cascade, roughly along a level curve of $\phi$, or directly in a
single jump) requires correspondingly small values of $p/q$: it must
squeeze around the peak into the valley along the $q$-axis. The
resulting flow is zonal with velocity along the $p$-axis, i.e.,
orthogonal to $\hat {\bm \gamma}$.

This conclusion is very general, following from a robust conservation
law \cite{foot:tcons} that operates under most situations considered in
the literature. Bounds on spectral energy transport imposed by
(\ref{18}) should help inform future more detailed flow computations.
Care, as well, should be taken in analyzing reduced models obtained
from (\ref{1}). The CHM equation \cite{CHM} possesses the extra
invariant, as well as enstrophy and the infinite potential vorticity
hierarchy. However some generalizations of this equation fail to do so,
at least in certain parameter ranges. Additional conservation might be
restored by re-including some neglected terms.


\begin{thebibliography}{}

\bibitem{DIIH2005} P. H. Diamond, S.-I. Itoh, K. Itoh, and T. S. Hahm,
    Plasma Phys.\ Control.\ Fusion \textbf{47}, R35 (2005).

\bibitem{DRHMFS1998} P. H. Diamond, M. N. Rosenbluth, F. L. Hinton,
    M. Malkov, J. Fleischer, and A. Smolyakov, in 17th IAEA Fusion
    Energy Conference, Yokohama, Japan (International Atomic Energy
    Agency, Vienna, 1998) IAEA–CN–69/TH3/1.

\bibitem{SDS2000} A. I. Smolyakov, P. H. Diamond, and V. I.
    Shevchenko, Phys.\ Plasmas \textbf{7}, 1349 (2000).

\bibitem{KWPSSS2005} T. D. Kaladze, D. J. Wu, O. A. Pokhotelov, R. Z.
    Sagdeev, L. Stenflo, and P. K. Shukla, Phys.\ Plasmas \textbf{12},
    122311 (2005).

\bibitem{KST2010} T. D. Kaladze, M. Shad, and L. V. Tsamalashvili,
    Phys.\ Plasmas \textbf{17}, 022304 (2010).

\bibitem{CNNQ2010} C. P. Connaughton, B. T. Nadiga, S. V. Nazarenko,
    and B. E. Quinn, J. Fluid Mech.\ \textbf{654}, 207 (2010)

\bibitem{Z1992} V. E. Zakharov, in \emph{Breaking Waves IUTAM
    Symposium,} Sydney, Australia, 1991, edited by M. L. Banner and R.
    H. J. Grimshaw (Springer, Berlin, 1992), pp.\ 69–-91.

\bibitem{CHM} J. G. Charney, Geophys.\ Publ.\ Oslo \textbf{17}, 1
    (1948); A. Hasegawa and K. Mima, Phys.\ Fluids \textbf{21}, 87
    (1978).

\bibitem{BNZ1991} A. M. Balk, S. V. Nazarenko, and V. E. Zakharov,
    Phys.\ Lett.\ A \textbf{152}, 276 (1991).

\bibitem{B1991} A. M. Balk, Phys.\ Lett.\ A \textbf{155}, 20 (1991).

\bibitem{BF1998} A. M. Balk and E. V. Ferapontov, in \emph{Nonlinear
    Waves and Weak Turbulence,} edited by V. E. Zakharov (American
    Mathematical Society, Translations Series 2, Providence, RI, 1998),
    Vol.\ \textbf{182,} pp.\ 1–-30.

\bibitem{BHW2011} A. M. Balk, F. van Heerden, and P. B. Weichman,
    Phys.\ Rev.\ E \textbf{83}, 046320 (2011).

\bibitem{NC1995} M. V. Nezlin and G. P. Chernikov, Plasma Phys.\
    Reports \textbf{21}, 922 (1995).

\bibitem{OPSSS2004} O. G. Onishchenko, O. A. Pokhotelov, R. Z. Sagdeev,
    P. K. Shukla, and L. Stenflo, Nonl.\ Processes Geophys.\
    \textbf{11}, 241 (2004).

\bibitem{MHM} Although much smaller in magnitude, the scalar
    nonlinearity is thought to be important because it greatly broadens
    the scale of wavenumbers that can lead to a modulational
    instability \cite{KWPSSS2005}.

\bibitem{foot:direction} Modern tokamaks (e.g., DIII-D, ITER) actually
    have non-circular poloidal cross-section, and the ``radial''
    direction already involves noncircular geometry.

\bibitem{ZS1988} V. E. Zakharov and E. I. Schulman, Physica D
    \textbf{29}, 283 (1988).

\bibitem{foot:Icorr} To be clear, once this condition is verified,
    it has indeed been proven that $I$ itself is conserved since the
    supplemental terms in (\ref{5}) may be oscillatory, but are of
    higher order in amplitude.

\bibitem{foot:driftwave} The drift wave mode is not visible directly in
    (\ref{3}), but may be projected out of the equations more directly
    using the refined field $s = {\cal Q} - \hat {\cal L}^{-1} {\bm
    \gamma} \cdot [(f/{\cal T}) {\bf \hat z} \times {\bf v} + \nabla
    h]$. This refinement basically serves to absorb the second
    term in (\ref{5}) into $I$. To linear order $s$ obeys the closed
    equation $\partial_t \hat {\cal L} s = -{\bm \gamma} \times \nabla
    s$, hence obeys dispersion relation (\ref{13}).

\bibitem{foot:GHM} GHM also fails to conserve enstrophy.

\bibitem{foot:approx} Note that in the context of the CHM
    equation, drift wave enstrophy and energy are conserved exactly,
    while the extra invariant remains approximate. However, in the
    context of the hydrodynamic equations all three are approximate due
    to interaction with non-drift-wave modes.

\bibitem{foot:tcons} By bounding the correction terms in ${\cal I},
    \dot {\cal I}$, it follows that $I$ can accumulate relative errors
    at most $O(\mu^2 \Omega t,\mu \epsilon^2 \Omega t)$ over time $t$:
    only for very large $\Omega t = O(1/\mu^2, 1/\mu \epsilon^2)$ may
    conservation be violated. This assumes all corrections add in
    phase; more likely, phases are random, leading to even longer
    conservation. Also, conservation is enhanced for zonal flows, and
    hence will further improve as ``condensation'' toward large scales
    proceeds.

\end{thebibliography}
\end{document}